\documentclass[
 reprint,
superscriptaddress,
reprint,
preprintnumbers,
nofootinbib,
 amsmath,amssymb,
 aps,
]{revtex4-2}

\usepackage{caption}
\usepackage{subcaption}

\usepackage{bbold}

\usepackage{amssymb}
\usepackage{amsfonts}
\usepackage{float}
\usepackage{braket, amsmath} 
\usepackage{cancel}
\usepackage{bbm}
\usepackage{graphicx}
\usepackage{dcolumn}
\usepackage{bm}
\usepackage[colorlinks=true, linkcolor=black, citecolor=black, urlcolor=blue]{hyperref}

\usepackage[utf8]{inputenc}

\usepackage{tikz} 

\usepackage{subfiles}

\begin{document}

\title{Entanglement dynamics through electromagnetic interactions in single-electron traps}

\author{Pablo Guillermo Carmona Rufo}
\affiliation{
Instituto de Física Teórica, UAM-CSIC, C/ Nicolás Cabrera 13-15, Campus de Cantoblanco, 28049 Madrid, Spain.} 

\author{Anupam Mazumdar}
\affiliation{Van Swinderen Institute for Particle Physics and Gravity,
University of Groningen, 9747AG, Groningen, The Netherlands.}

\author{Carlos Sabín}
\affiliation{Departamento de Física Teórica and CIAFF, Universidad Autónoma de Madrid, 28049, Madrid, Spain.}

\begin{abstract}
We study the dynamics of quantum entanglement between two harmonically trapped electrons interacting via the electromagnetic force. Starting from two-mode Gaussian states at thermal equilibrium, we make use of the covariance matrix formalism in order to compute the logarithmic negativity of the evolved state as a quantitative measure of entanglement. We analyze two initial configurations: thermal single-mode and two-mode squeezed states, and describe the time evolution of entanglement in the system for different values of squeezing and temperature, while identifying the parameter regimes accessible to current and near-future single-electron trap experiments.

\end{abstract}
\maketitle

\section{Introduction}
The trapping and control of quantum systems has unquestionably emerged as one of the most successful areas of quantum studies \cite{Anupam1,start1,start2,start3,start4}. Charged particles can be trapped through the use of time dependent electric fields in Paul traps \cite{Paul1,Paul2}, avoiding Earnshaw's theorem of electromagnetism \cite{Earnshaw}. These setups have already been employed in a wide range of applications, such as precision measurements \cite{Precision} and metrology \cite{Metrology}. The trapping of charged particles opens the door to a variety of emerging applications including tests of fundamental physics \cite{fundamental1,fundamental2,fundamental3} and the development of scalable quantum computers \cite{qc1,qc2,qc3,qc4,qc5}. Specifically, single-electron traps provide the possibility of manipulating the particle's spin as a resource for quantum information processing \cite{qi1,qi2,qi3,qi4,qi5,qi6}.

The trapping of electrons by maintaining them in a confined region by means of modulated electromagnetic fields has already been explored by several groups \cite{qi4,qi5,groups1,groups2,groups3,groups4}. This type of experiments is of particular interest, since the possibility of controlling the motional and spin state of the electrons provides a framework for the investigation of diverse quantum phenomena. For instance, it has been shown that electromagnetic interactions between quantum systems tend to generate quantum entanglement between them \cite{Anupam1,Yang2015}. Moreover, they are analogous to gravitational interactions, where recent works have heavily focused on the study of entanglement in table-top experiments with the goal of probing fundamental physics aspects \cite{Anupam1,qgem1,qgem2,MatterLight1,MatterLight2,darkmatter,weak2,Massivegraviton,string}. 

In this work, we investigate the presence of quantum entanglement in the interaction between two harmonically trapped electrons, considering our system to be in two different types of squeezed thermal quantum states prior to the interaction. We analyze the role that the different physical parameters of the system play in the generation and detection of entanglement, including the squeezing coefficient and the average number of quanta per mode.

For this purpose, we consider a set of two particles trapped in adjacent harmonic traps. In order to study their system's properties and evolution, we make use of Gaussian states theory. Gaussian states naturally appear in the description of ion traps, as well as plenty of other physical models, such as optomechanical or nanomechanical oscillators and gases of cold atoms \cite{Olivares2012}. We study the time evolution of our system, providing a solution to its equations of motion, and then analyze its properties through its covariance matrix, which fully characterizes it. To compute the transformation of the covariance matrix, we will employ the covariance matrix formalism \cite{Covariance1,Covariance2}, which facilitates the study of entanglement through the use of symplectic eigenvalues and local symplectic invariants. This allows us to study the evolution of entanglement in the interaction by means of the logarithmic negativity. 
We consider two different scenarios. First, the initial two-mode state is separable -thermal single-mode squeezed state-, and we find that the interaction generates entanglement after an interaction time depending on the temperature and the amount of initial squeezing. Then we consider an initial thermal two-mode squeezing state so that entanglement is already present and what we find is the degradation of quantum correlations with the interaction,
depending again on temperature and squezing.

The structure of this paper is the following: In Section \ref{sec:model}, we introduce the theoretical model of two single-particle traps.  In Section \ref{sec:eom} we solve the system's equations of motion, while in Section \ref{sec:cmm}, we give a more detailed explanation of the covariance matrix formalism. In Section \ref{sec:results}, we showcase our results for the logarithmic negativity of the system in the two different scenarios described above. Finally, we discuss these results in Section \ref{sec:discussion} and show the analytical expressions that we have computed and used in our results in Appendix A and B.
\section{Model}
\label{sec:model}
Our model \cite{Anupam1} has two particles trapped in harmonic traps on the x-axis, at positions:
\begin{equation}
    \hat{r}_1=-d/2+\hat{x}_1~~~~~\hat{r}_2=d/2+\hat{x}_2,
\end{equation}
where $\hat{x}_1$ and $\hat{x}_2$ are the quantum fluctuations around the equilibrium positions. We consider that each particle has mass $m$ and charge $-e$. Plus, we assume that the frequency of both modes is equal and represented by $\omega$. The Hamiltonian of the system will therefore be given by:
\begin{equation}
    \hat{H}=\sum_{i=1,2}(\frac{\hat{p}^2_i}{2m}+\frac{1}{2}m\omega^2\hat{x}_i^2)+\hat{H}_\text{int}.
\end{equation}
For our study, we are going to consider that the interaction Hamiltonian $\hat{H}_\text{int}$ is composed of the electromagnetic interactions up to first order in the speed of light $c$. Specifically:
\begin{equation}
    \hat{H}_\text{int}=\hat{H}_C+\hat{H}_D,
\end{equation}
where $\hat{H}_C$ is the usual Coulomb potential and $\hat{H}_D$ is known as the Darwin Hamiltonian, which contains the momentum contributions that arise at the leading order relativistic correction. More details on this can be found on \cite{Anupam1}. The form of these two terms is given by:
\begin{subequations}
\begin{align}
    \hat{H}_C&=\frac{e^2}{4\pi\varepsilon_0|\hat{r}|}, \label{eq:hc} \\
    \hat{H}_D&=-\frac{e^2}{8\pi\varepsilon_0m^2c^2}\left[\frac{\left(\hat{p}_1\cdot\hat{p}_2\right)}{|\hat{r}|}+\frac{\left(\hat{p}_1\cdot \hat{r}\right)\left(\hat{p}_2\cdot \hat{r}\right)}{|\hat{r}|^3}\right], \label{eq:hd}
\end{align}
\end{subequations}
where $\hat{r}$ is the distance between the two particles, given by:
\begin{equation}
    \hat{r}=\hat{r}_2-\hat{r}_1=d+\hat{x}_2-\hat{x}_1.
\label{eq:distance}
\end{equation}
The derivation of the terms in (\ref{eq:hc}) and (\ref{eq:hd}) can be found in \cite{hc3}, with alternative derivations in \cite{hc1,hc2}. Now, we can rewrite the quantum operators using the ladder operators in the following way:
\begin{equation}
    \hat{x}_j=\delta x\left(\hat{a}^\dagger_i+\hat{a}_i\right)~~~~~\hat{p}_j=i\delta p\left(\hat{a}^\dagger_i-\hat{a}_i\right),
\label{eq:operators}
\end{equation}
where:
\begin{equation}
    \delta x=\sqrt{\frac{\hbar}{2m\omega}}~~~~~
    \delta p=\sqrt{\frac{m\omega\hbar}{2}}.
\end{equation}
We can now substitute (\ref{eq:distance}) and (\ref{eq:operators}) in Equations (\ref{eq:hc}) and (\ref{eq:hd}) and expand in $\hat{x}_1$ and $\hat{x}_2$ to obtain the leading order interaction Hamiltonian, assuming the distance $d$ between the two particles to be much larger than the fluctuations $\hat{x}_i$. This results in:
\begin{multline}
     \hat{H}_\text{int}=\hbar\left(g_C-g_D\right)(\hat{a}^\dagger_1\hat{a}_2+\hat{a}_1\hat{a}^\dagger_2)\\+\hbar\left(g_C+g_D\right)(\hat{a}_1\hat{a}_2+\hat{a}^\dagger_1\hat{a}^\dagger_2),
\end{multline}
where we have defined the following couplings that emerge from the Coulomb and Darwin Hamiltonians respectively:
\begin{equation}
    g_C\equiv-\frac{e^2}{4\pi\varepsilon_0m\omega d^3}~~~~~g_D\equiv\frac{3\omega e^2}{16\pi\varepsilon_0mdc^2}.
\end{equation}
A discussion on higher order terms can be found in \cite{Anupam1}.

\section{Equations of motion}
\label{sec:eom}

In order to study the time evolution of our system, we solve the dynamics by calculating the time evolution of the ladder operators using the method given in \cite{Canosa_2015}. The Heisenberg equations of motion for said operators will be:
\begin{equation}
    i\hbar\dot{\hat{a}}_j=[\hat{a}_j,\hat{H}]~~~j=1,2,
\label{eq:commutators}
\end{equation}
which can be written in matrix form as:
\begin{equation}
    i\begin{pmatrix}\dot{\boldsymbol{\hat{a}}}\\\dot{\boldsymbol{\hat{a}}}^\dagger\end{pmatrix}=\mathcal{H}\begin{pmatrix}\boldsymbol{\hat{a}}\\\boldsymbol{\hat{a}}^\dagger\end{pmatrix},
\end{equation}
where:
\begin{equation}
\boldsymbol{\hat{a}}=\begin{pmatrix}\hat{a}_1\\\hat{a}_2\end{pmatrix}~~~~~\boldsymbol{\hat{a}}^\dagger=\begin{pmatrix}\hat{a}_1^\dagger\\\hat{a}_2^\dagger\end{pmatrix}
\end{equation}
and
\begin{equation}
    \mathcal{H}=\begin{pmatrix}
\omega & g_C - g_D & 0 & g_C + g_D \\
g_C - g_D & \omega & g_C + g_D & 0 \\
0 & -(g_C + g_D) & -\omega & -(g_C - g_D) \\
-(g_C + g_D) & 0 & -(g_C - g_D) & -\omega
\end{pmatrix},
\end{equation}
which is obtained calculating the commutators in (\ref{eq:commutators}), which are shown in  Appendix A for the sake of completeness. With this, we can calculate the exact solution of this equation as:
\begin{equation}
    \begin{pmatrix}\boldsymbol{a}(t)\\\boldsymbol{a}^\dagger(t)\end{pmatrix}=\mathcal{U}(t)\begin{pmatrix}\boldsymbol{a}(0)\\\boldsymbol{a}^\dagger(0)\end{pmatrix},
\end{equation}
where
\begin{equation}
    \mathcal{U}(t)=\exp{[-i\mathcal{H}t]},
\label{eq:Ufull}
\end{equation}
which can be computed through the diagonalization of $\mathcal{H}$. The result for $\mathcal{U}$ is displayed in Appendix A. This expression for our time evolved operators can be understood as a proper time-dependent Bogoliubov transformation of the form \cite{Covariance1,Covariance2}:
\begin{equation}
    a_m=\sum_n\left(\alpha^*_{mn}a_n+\beta^*_{mn}a^\dagger_n\right),
\end{equation}
a property that will help us explore the system through Gaussian state theory in the following Section.

\section{Covariance Matrix Formalism}
\label{sec:cmm}

We now consider the covariance matrix formalism, which has been previously used in multiple works so as to analyze entanglement in quantum field theory \cite{Friis01012013,Adesso_2007,Adesso_2012}, since it is applicable to systems that consist of any number of bosonic modes. For Gaussian states of a bosonic field, all the relevant information about the state is encoded in the first and second moments of the field. Particularly, the second moments are described by the quadrature vector, which, for a two-mode state, is given by:
\begin{equation}
    \mathbf{Y}\equiv(\hat{x}_1,\hat{p}_1,\hat{x}_2,\hat{p}_2),
\end{equation}
and the covariance matrix, of the form:
\begin{equation}
\sigma_{ij}=\braket{\{Y_i,Y_j\}}-2\braket{Y_i}\braket{Y_j}.
\end{equation}
This formalism allows for simplified calculations and has been proven to be useful to compute measure of multipartite entanglement for Gaussian states \cite{Illuminati,Adesso_2012}.

Every unitary transformation generated by a quadratic Hamiltonian can be represented as a symplectic matrix $S$ in phase space. Said transformations form the real symplectic group, which is the group of real matrices that leave the symplectic form $\Omega$ invariant, that is, $S\Omega S^T=\Omega$, where $\Omega=\oplus_{i=1}^n \Omega_i$ and: $\Omega_i=\begin{pmatrix}0 & 1 \\-1 & 0\end{pmatrix}$.
The Bogoliubov transformation associated to the time evolution of our system can be translated into the covariance matrix language. For a two-mode system like the one we have, the symplectic matrix for the transformation in can be written in terms of the Bogoliubov coefficients $(\alpha_{mn}$, $\beta_{mn})$ as:
\begin{equation}
S=\begin{pmatrix}\mathcal{M}_{11}&\mathcal{M}_{12}\\\mathcal{M}_{21}&\mathcal{M}_{22}\end{pmatrix},
\end{equation}
where
\begin{equation}
\mathcal{M}_{mn}=\begin{pmatrix}\Re\left(\alpha_{mn}-\beta_{mn}\right)&\Im\left(\alpha_{mn}+\beta_{mn}\right)\\ -\Im\left(\alpha_{mn}-\beta_{mn}\right)&\Re\left(\alpha_{mn}+\beta_{mn}\right)\end{pmatrix},
\label{eq:Mcomps1}
\end{equation}
where $\Re$ and $\Im$ represent the real imaginary parts, respectively. We now analyze what our covariance matrix looks like before and after the transformation. In this case, the initial covariance matrix will be given by:
\begin{equation}
    \sigma=\begin{pmatrix}\psi_1&\phi_{12}\\\phi^T_{12}&\psi_{2}\end{pmatrix},
\end{equation}
where $\psi_1$ and $\psi_2$ are the reduced covariance matrices of modes 1 and 2, respectively and $\phi_{12}$ is a $2\times2$ matrix accounting for the correlations between the modes.
Then, the components of the transformed covariance matrix, which fulfills $\tilde{\sigma}=S\sigma S^T$, will be:
\begin{equation}
    \tilde{\sigma}=\begin{pmatrix}C_{11}&C_{12}\\C_{21}&C_{22}\end{pmatrix},
\label{eq:transformed}
\end{equation}
where
\begin{multline}
    C_{ij}=\mathcal{M}^T_{1i}\psi_1\mathcal{M}_{1j}+\mathcal{M}^T_{2i}\phi^T_{12}\mathcal{M}_{1j}\\+\mathcal{M}^T_{1i}\phi^T_{12}\mathcal{M}_{2j}+\mathcal{M}^T_{2i}\psi_{2}\mathcal{M}_{2j}.\label{eq:cs}
\end{multline}
 We want to study the entanglement generated by the interaction within the covariance matrix formalism. For pure systems, we could achieve this by first calculating the symplectic eigenvalue $\sigma^{(1)}(t)$ as $\sigma^{(1)}(t)=\sqrt{\det{\boldsymbol{\sigma}_1}}$, where $\boldsymbol{\sigma}_1$ is the reduced covariance matrix, simply given by $C_{11}$, and then making use of the von Neumann entanglement entropy $S_\text{ent}(\rho)$ \cite{Demarie_2018,Olivares2012,Kumar2023continuousvariable}. However, we will not be dealing with pure states, since, in order to account for environmental noise and provide a more realistic description of experimentally accessible systems, we will consider our system to be in a multi-mode Gaussian state at thermal equilibrium at temperature $T$, with an average number $N_k$ of quanta in the $k$-th mode given by \cite{Olivares2012}:
\begin{equation}\label{eq:thermquant}
    N_k=\left(e^{\frac{\hbar\omega}{k_BT}}-1\right),
\end{equation}
with $k_B$ being the Boltzmann constant. In this case, we can no longer calculate the entanglement using the reduced covariance matrix, instead, we need to calculate the smallest symplectic eigenvalue of the partial transpose of the transformed covariance matrix in (\ref{eq:transformed}). For this, following \cite{Olivares2012}, we must define a series of quantities that are left unchanged by a series of local symplectic transformations:
\begin{multline}
\tilde{I}_1=\det[C_{11}]~~~\tilde{I}_2=\det[C_{22}]~~~\tilde{I}_3=\det[C_{12}]\\
\tilde{I}_4=\det[\tilde{\sigma}]~~~\tilde{\Delta}(\tilde{\sigma})=I_1+I_2-2I_3.
\label{eq:invariants}
\end{multline}
With this, one can calculate the symplectic eigenvalues of the partially transposed covariance matrix as:
\begin{equation}
    \tilde{d}_{\pm}=\sqrt{\frac{\tilde{\Delta}(\tilde{\sigma})\pm\sqrt{\tilde{\Delta}(\tilde{\sigma})^2-4\tilde{I}_4}
    }{2}}.
\label{eq:dminus}
\end{equation}
We can then use the smallest eigenvalue $\tilde{d}_-$ to calculate the logarithmic negativity of the system as:
\begin{equation}
    E_\mathcal{N}=\max\{0,-\log{\tilde{d}_-}\},
 \label{eq:negativity}
\end{equation}
which is a simple increasing monotone function of the minimum symplectic eigenvalue, making it a proper candidate for the quantitative evaluation of entanglement.



\section{Results}
\label{sec:results}

We now analyze the interaction using squeezed states \cite{Anupam1}, which enable control over the initial delocalization in position and momentum in the context of Gaussian states \cite{Weedbrook,Illuminati2}. We will investigate two different scenarios, one where our system is composed of two single-mode squeezed states and one with two-mode squeezing between them. The single-mode and two-mode squeezing operators acting on the system modes are given by
\begin{equation}
    S_1(\xi)=\exp\{\frac12\left[\xi \hat{a}^{\dagger2}-\xi^*\hat{a}^2\right]\}
\end{equation}
and
\begin{equation}
    S_2(\xi)=\exp\{\xi \hat{a}_1^\dagger\hat{a}_2^\dagger-\xi^*\hat{a}_1\hat{a}_2\}
\end{equation}
respectively, where $\xi$ is the squeezing coefficient. The numerical values for $\xi$ will be constrained by experimental limitations. In particular, the experience in quantum optical systems shows that it is very difficult to attain high values of the squeezing parameter, with $\xi=1.73$ being the highest reported in LIGO \cite{LIGO}.

For the single-mode squeezing case, each electron state can be squeezed locally within its own trap, independently of the electromagnetic interaction between the traps. However, for the two-mode squeezed state, since generating the initial correlations between the two electrons requires some interaction between them, it is worth remarking that, for the sake of simplicity, we treat the preparation of the initial state as an idealized process occurring prior to $t=0$, namely, the moment where the separated traps are created and the interaction begins. 

\subsection{Single-mode squeezing}

The covariance matrix for a thermal two-mode single-mode squeezed state is given by \cite{Olivares2012}:
\begin{equation}
    \psi_1=\psi_{2}=\begin{pmatrix}\sigma_{11} & \sigma_{12}\\ \sigma_{21} & \sigma_{22}\end{pmatrix}~~~\phi_{12}=0,
\label{eq:psisinglemode}
\end{equation}
where the elements are:
\begin{subequations}
\begin{align}
    \sigma_{ii}=(1+2N)\left[\cosh{(2r)}-(-1)^i\sinh{(2r)}\cos{(\theta)}\right],\label{eq:sigmasinglea}\\
    \sigma_{12}=\sigma_{21}=(1+2N)\sinh{(2r)}\sin{(\theta)},\label{eq:sigmasingleb}
\end{align}
\end{subequations}
with $\xi=re^{i\theta}$ and $N=N_1=N_2$ is given by Eq. (\ref{eq:thermquant}). Therefore, we have:
\begin{equation}
    C_{ij}=\mathcal{M}^T_{1i}\psi_1\mathcal{M}_{1j}+\mathcal{M}^T_{2i}\psi_{2}\mathcal{M}_{2j},
\end{equation}
which we can calculate using (\ref{eq:Mcomps1}) and (\ref{eq:psisinglemode}). Then, we can work out the smallest symplectic eigenvalue, which we will refer to as $\tilde{d}_{\text{SM}}$, with (\ref{eq:invariants}) and (\ref{eq:dminus}), as well as the logarithmic negativity using (\ref{eq:negativity}). 
In order to perform our calculations, we have defined a few adimensional quantities:

\begin{equation}
    \tilde{g}_C=\frac{g_C}{\omega}~~~\tilde{g}_D=\frac{g_D}{\omega}~~~
    \alpha=\omega t 
\end{equation}

\begin{figure*}[t]
    \centering

    \begin{subfigure}{0.45\textwidth}
        \centering
        \includegraphics[width=\linewidth]{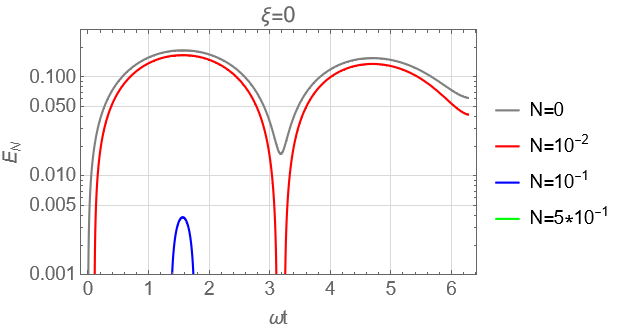}
        \caption{}
    \end{subfigure}

    \par\medskip

    \begin{subfigure}{0.45\textwidth}
        \centering
        \includegraphics[width=\linewidth]{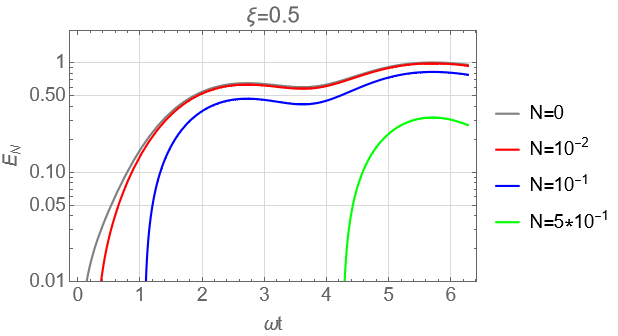}
        \caption{}
    \end{subfigure}
    \hfill
    \begin{subfigure}{0.45\textwidth}
        \centering
        \includegraphics[width=\linewidth]{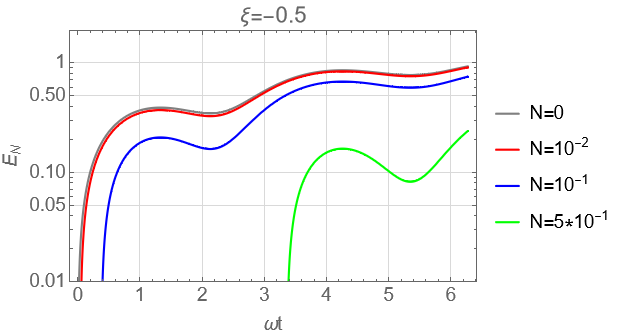}
        \caption{}
    \end{subfigure}

    \par\medskip

    \begin{subfigure}{0.45\textwidth}
        \centering
        \includegraphics[width=\linewidth]{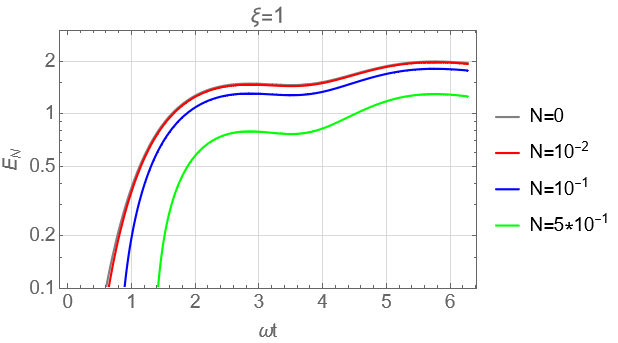}
        \caption{}
    \end{subfigure}
    \hfill
    \begin{subfigure}{0.45\textwidth}
        \centering
        \includegraphics[width=\linewidth]{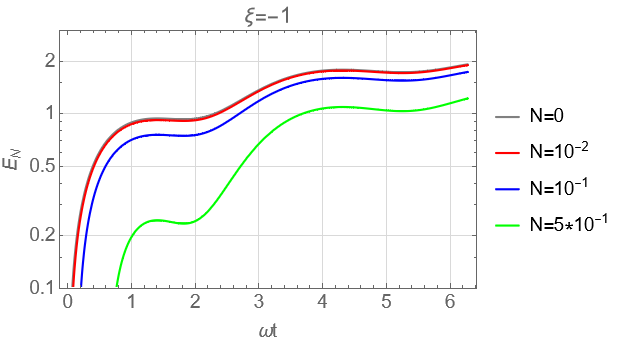}
        \caption{}
    \end{subfigure}
    \caption{Values of the logarithmic negativity of the single-mode squeezed thermal state plotted for $\alpha=\omega t \in[0,2\pi]$  and for (a) $\xi=0$, (b) $\xi=0.5$, (c) $\xi=-0.5$, (d) $\xi=1$ and (e) $\xi=-1$. The values used for the physical parameters of the system are  $\omega=10$ GHz and $d=3$ $\mu$m. In each plot, each curve is calculated for a different value of the thermal number, which depend on experimental constraints as well. A frequency of $\omega=10$ GHz and the temperature that has been able to be reached in recent experiments, which ranges around $50$ to $100$ mK \cite{Thermal1,Yu2025}, result in a thermal number of $N\sim0.1$. Thus, the curves with values for $N$ lower than this threshold represent what possible future experimental improvements might achieve.}
    \label{fig:Sm}
\end{figure*}

Therefore, the full expression for the eigenvalue depends on six parameters: $\tilde{g}_C$, $\tilde{g}_D$, $\alpha$, $N$, $r$ and $\theta$, with the dependence on the distance between the electrons being contained in  $\tilde{g}_C$ and $\tilde{g}_D$.
Even though the exact formula is rather lengthy, we can make some approximations in order to be able to display an analytical result. First, with the values picked for the parameters, we notice that $\tilde{g}_D\ll\tilde{g}_C$, meaning that the Darwin term of the electromagnetic interaction is negligible for our analysis with respect to the leading order Coulomb term. Therefore, we will showcase the corresponding formulas with $\tilde{g}_D=0$. Plus, we consider only two cases: $\theta=0$ and $\theta=\pi$, that is, when the squeezing coefficient is a real number. The expressions for $\tilde{d}_{\text{SM}}$ in these cases can be found in Appendix B, while the results for the logarithmic negativity are shown in Figure \ref{fig:Sm}, for different values of the squeezing and the average number of quanta in each mode $N$. The results in the figures match with what we would expect: The system presents no entanglement at $t=0$, but the interaction entangles the electrons' states as time passes, depending on the squeezing and the noise. If we first analyze the $\xi=0$ case, the curve with $N=0.5$ is not able to generate entanglement. In fact, in this case, entanglement is only present throughout a whole time evolution cycle, that is, between $\alpha=0$ and $\alpha=2\pi$, for the $N=0$ case out of the ones considered. The $N=0.01$ curve presents sudden death and revival of entanglement \cite{death0,death1,death2,death3,deathions} around $\omega t=\pi$, while for the $N=0.1$ one the entanglement only shows up for a brief part of the cycle, around $\omega t=\pi/2$. For $|\xi|=0.5$ and $|\xi|=1$, the presence of the squeezing diminishes the dip in entanglement around $\omega t=\pi$. Moreover, the growth of entanglement with time is clear, and, as can be seen, the larger the thermal number $N$ we consider, the longer it takes for it to be detected. It is worth noting that the pairs of figures with the same absolute value of the squeezing parameter are not fully identical, meaning that the entanglement evolution presents a dependence on the angle $\theta$, as implied in (\ref{eq:da})-(\ref{eq:db}). For the $N=0.5$ curve, entanglement appears around $\omega t=1$ for both $|\xi|=1$ curves, but this difference becomes particularly noticeable for $|\xi|=0.5$, where it takes relatively longer for entanglement to arise when the squeezing is positive.

\subsection{Two-mode squeezing}
\begin{figure*}[t]
    \centering

    \begin{subfigure}{0.45\textwidth}
        \includegraphics[width=
\linewidth]{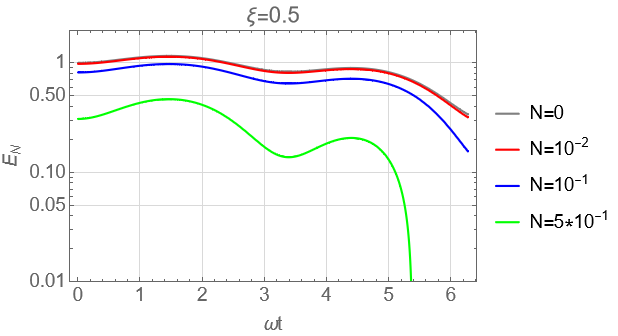}
        \centering
        \caption{}
    \end{subfigure}
    \hfill
    \begin{subfigure}{0.45\textwidth}
        \includegraphics[width=
\linewidth]{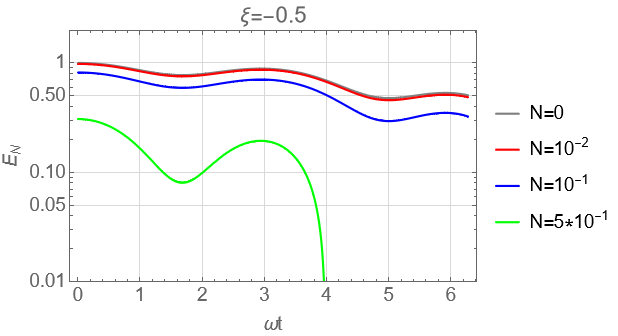}
        \centering
        \caption{}
    \end{subfigure}
    \hfill
    \begin{subfigure}{0.45\textwidth}
    \includegraphics[width=
\linewidth]{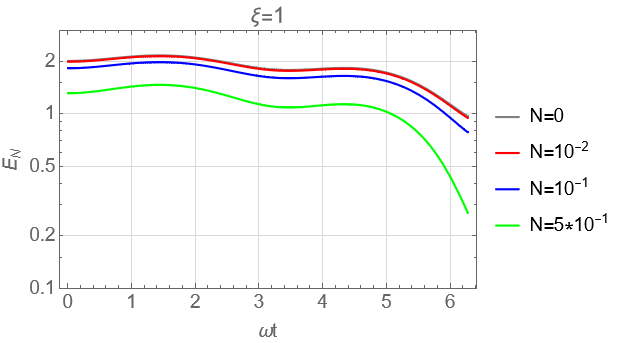}
        \centering

        \caption{}
    \end{subfigure}
    \hfill
    \begin{subfigure}{0.45\textwidth}
    \includegraphics[width=
\linewidth]{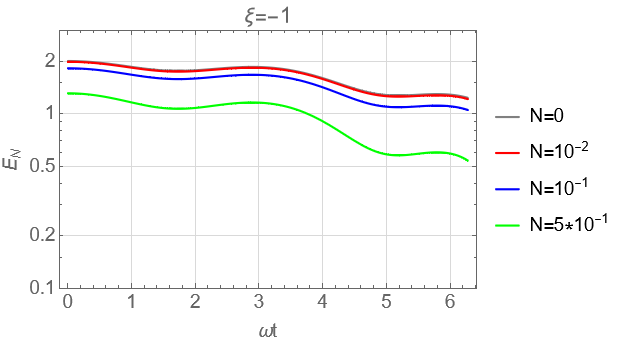}
        \centering

        \caption{}
    \end{subfigure}
    \caption{Values of the logarithmic negativity of the two-mode squeezed thermal state plotted for (a) $\xi=0.5$, (b) $\xi=-0.5$, (c) $\xi=1$ and (d) $\xi=-1$. The discussion around the physical parameters is fully equivalent to the one for the single-mode case, detailed in the footnote of Figure \ref{fig:Sm}. }
    \label{fig:tm}
\end{figure*}

 In this case, the covariance matrix for a two-mode squeezed thermal state will be \cite{Olivares2012}:
\begin{equation}
    \psi_1=A\mathbb{1}_2,~~~
    \psi_{2}=B\mathbb{1}_2,~~~
    \phi_{12}=C\mathbf{R}_\theta,~~~
\end{equation}
where
\begin{equation}
    \mathbf{R}_\theta=\begin{pmatrix}\cos{\theta} & \sin{\theta}\\ \sin{\theta} & -\cos{\theta}\end{pmatrix}
\end{equation}
and:
\begin{subequations}
\begin{align}
    A&=(1+N_1+N_2)\cosh{\left(2r\right)}+(N_1+N_2),\label{eq:Aa}\\
    B&=(1+N_1+N_2)\cosh{\left(2r\right)}-(N_1-N_2),\label{eq:Bb}\\
    C&=(1+N_1+N_2)\sinh{\left(2r\right)}.\label{eq:Cc}
\end{align}
\end{subequations}
If we again assume both modes to have the same thermal number $N_1=N_2\equiv N$, this simply reduces to:
\begin{subequations}\label{eq:psitwomode}
\begin{align}
    \psi_1&=\psi_{2}=(1+2N)\cosh{\left(2r\right)}\mathbb{1}_2,\label{eq:psitwomodea}\\
    \phi_{12}&=(1+2N)\sinh{\left(2r\right)}\mathbf{R}_\theta.\label{eq:psitwomodeb}
\end{align}
\end{subequations}
We now need the full formula (\ref{eq:cs}), which 
we can calculate using (\ref{eq:Mcomps1}) and (\ref{eq:psitwomode}), and then the smallest symplectic eigenvalue $\tilde{d}_{\text{TM}}$ in the same way as the previous case. After carrying out the same approximations as for the single-mode squeezing case, the expressions for $\tilde{d}_{\text{TM}}$ are also in Appendix B. We show the results for the logarithmic negativity in Figure \ref{fig:tm}, again for different values of the squeezing and the thermal occupation number $N$. We do not show the $\xi=0$ case this time, since it would look identical to Figure \ref{fig:Sm}a. For $\xi\neq0$ however, in this case, due to the initial two-mode squeezing, the states are already entangled at $t=0$ and the amount of entanglement between them decreases as the system evolves. Once again, the way the entanglement evolves depends on the sign of the squeezing coefficient, which can be noticed in (\ref{eq:dc})-(\ref{eq:dd}) too. As the results in Figures \ref{fig:tm}a and \ref{fig:tm}b illustrate, this difference is quite evident for $|\xi|=0.5$, where the entanglement dies considerably earlier for $N=0.5$ and negative squeezing.

\section{Discussion}
\label{sec:discussion}

In this article, we explored the evolution of entanglement in a system composed of two single-electron harmonic traps. In order to achieve this, we first were able to study the system's time evolution by providing a full analytical solution to the system's equations of motion, and then we studied our system through the covariance matrix formalism, using the logarithmic negativity as an entanglement measure. Our results show that the dynamics of the entanglement depends on the type of squeezing we consider: For two single-mode squeezed initial states, the electrons begin in unentangled states, and become entangled during the evolution, with the degree of entanglement depending on the value of the squeezing parameter and the thermal occupation number, while an initial two-mode squeezing already exhibits entanglement between the electrons prior to the start of the interaction and is degraded by it. Plus, in the considered parameter regime, we notice that a sudden death and revival of entanglement might emerge when we assume an unsqueezed initial state.

Beyond the bipartite scenario considered in this work, an interesting direction for future research is the extension of the present framework to systems comprising three electron traps. Extending the analysis to this regime would enable the study of tripartite entanglement and other uniquely multipartite quantum phenomena. Another future idea that could be of interest to consider could be the development of an entanglement witness tailored to the electron trap system considered here.

\section*{Acknowledgements}
P.G.C.R. acknowledges:  Grant PRE2022-102488 funded by MCIN/AEI/10.13039/501100011033 and FSE+, with project code: PID2021-127726NB-I00. A.M.’s research is funded in part by the Gordon and Betty Moore Foundation through Grant GBMF12328, DOI 10.37807/GBMF12328. C.S. acknowledges financial support through the Ramón y Cajal Programme (RYC2019-028014-I) and Consolidación Investigadora (CNS2024-154149). 

\subsection*{Appendix A: Analytical solution of the time evolution operator}

\label{app:appendixA}

The result for the commutators in (\ref{eq:commutators}) is:
\begin{widetext}
\begin{subequations}
\begin{align}
        [\hat{a}_1,\hat{H}]=\hbar\omega \hat{a}_1+\hbar(g_C-g_D)\hat{a}_2+\hbar(g_C+g_D)\hat{a}^\dagger_2\\
        [\hat{a}_2,\hat{H}]=\hbar\omega \hat{a}_2+\hbar(g_C-g_D)\hat{a}_1+\hbar(g_C+g_D)\hat{a}^\dagger_1
\end{align}
\end{subequations}
\end{widetext}
This allows us to now provide a full analytical solution for the time evolution operator $\mathcal{U}$, following (\ref{eq:Ufull}) in the main text:
\begin{widetext}
\begin{subequations}\label{eq:U}
\begin{align}
    \mathcal{U}_{11}&=\mathcal{U}_{22}=\mathcal{U}_{33}^*=\mathcal{U}_{44}^*=\frac12\left[\cos{\left(\Omega_+t\right)}+\cos{\left(\Omega_-t\right)}-\left(\frac{\sin{\left(\Omega_+t\right)}}{\Omega_+}(\omega+g_C-g_D)+\frac{\sin{\left(\Omega_-t\right)}}{\Omega_-}(\omega-g_C+g_D)\right) i\right]\label{eq:Ua}\\
    \mathcal{U}_{12}&=\mathcal{U}_{21}=\mathcal{U}_{34}^*=\mathcal{U}_{43}^*=\frac12\left[\cos{\left(\Omega_+t\right)}-\cos{\left(\Omega_-t\right)}-\left(\frac{\sin{\left(\Omega_+t\right)}}{\Omega_+}(\omega+g_C-g_D)-\frac{\sin{\left(\Omega_-t\right)}}{\Omega_-}(\omega-g_C+
    g_D)\right) i\right]\label{eq:Ub}\\
    \mathcal{U}_{13}&=\mathcal{U}_{24}=\mathcal{U}_{31}^*=\mathcal{U}_{42}^*=\frac{\left(g_C+g_D\right)\left(-\Omega_-\sin{\left(\Omega_+t\right)}+\Omega_+\sin{\left(\Omega_-t\right)}\right)}{2\Omega_+\Omega_-}i\label{eq:Uc}\\
    \mathcal{U}_{14}&=\mathcal{U}_{23}=\mathcal{U}_{32}^*=\mathcal{U}_{41}^*=-\frac{\left(g_C+g_D\right)\left(\Omega_-\sin{\left(\Omega_+t\right)}+\Omega_+\sin{\left(\Omega_-t\right)}\right)}{2\Omega_+\Omega_-}i\label{eq:Ud}
\end{align}
\end{subequations}
\end{widetext}
Here, we have defined $\Omega_+=\sqrt{\omega+2g_C}\sqrt{\omega-2g_D}$ and $\Omega_-=\sqrt{\omega-2g_C}\sqrt{\omega+2g_D}$. The computed Bogoliubov coefficients are therefore:
\begin{widetext}
\begin{subequations}
\begin{align}
    \alpha_{11}&=\alpha_{22}=\mathcal{U}_{11}^*\label{eq:Boga}\\
    \alpha_{12}&=\alpha_{21}=\mathcal{U}_{12}^*\label{eq:Bogb}\\
    \beta_{11}&=\beta_{22}=-\mathcal{U}_{13}^*=\mathcal{U}_{13}\label{eq:Bogc}\\
    \beta_{12}&=\beta_{21}=-\mathcal{U}_{14}^*=\mathcal{U}_{14}\label{eq:Bogd}
\end{align}
\end{subequations}
\end{widetext}
and the obtained result for the $\mathcal{M}_{ij}$ components using (\ref{eq:Mcomps1}) is:
\begin{widetext}
\begin{subequations}\label{eq:Emesapp}
\begin{align}
    \mathcal{M}_{kk}&=\mathcal{M}_{k'k'}=\frac12\begin{pmatrix}\cos{\left(\Omega_+t\right)}+\cos{\left(\Omega_-t\right)} & \frac{\sin{\left(\Omega_+t\right)}\left(\omega-2g_D\right)}{\Omega_+}+\frac{\sin{\left(\Omega_-t\right)}\left(\omega+2g_D\right)}{\Omega_-}\\ -\frac{\sin{\left(\Omega_+t\right)}\left(\omega+2g_C\right)}{\Omega_+}-\frac{\sin{\left(\Omega_-t\right)}\left(\omega-2g_C\right)}{\Omega_-} & \cos{\left(\Omega_+t\right)}+\cos{\left(\Omega_-t\right)}\end{pmatrix}\label{eq:Ma}\\
   \mathcal{M}_{kk'}&=\mathcal{M}_{k'k}=\frac12\begin{pmatrix}\cos{\left(\Omega_+t\right)}-\cos{\left(\Omega_-t\right)} & \frac{\sin{\left(\Omega_+t\right)}\left(\omega-2g_D\right)}{\Omega_+}-\frac{\sin{\left(\Omega_-t\right)}\left(\omega+2g_D\right)}{\Omega_-}\\ -\frac{\sin{\left(\Omega_+t\right)}\left(\omega+2g_C\right)}{\Omega_+}+\frac{\sin{\left(\Omega_-t\right)}\left(\omega-2g_C\right)}{\Omega_-} & \cos{\left(\Omega_+t\right)}-\cos{\left(\Omega_-t\right)}\end{pmatrix}\label{eq:Mb}
\end{align}
\end{subequations}
\end{widetext}

\subsection*{Appendix B: Analytical solution of the smallest symplectic eigenvalues}

\label{app:appendixB}

With all the results of the previous section, the covariance matrix for the single-mode squeezed state and two-mode squeezed states can be computed as explained in the main text. From them, we can use (\ref{eq:dminus}) to 
calculate the smallest symplectic eigenvalue for both the single-mode and two-mode squeezing cases. We give here the expressions for $\tilde{g}_D=\theta=0$ and $\tilde{g}_D=0$, $\theta=\pi$:
\begin{widetext}

\begin{equation}
\begin{aligned}
&\tilde{d}_{\text{SM}}(\tilde{g}_C,\tilde{g}_D=0,\alpha,N,r,\theta=0)=
\\[1ex]
&\frac{1}{2}(1+2N)
\sqrt{
\Biggl(
\frac{1}{\lambda_-^2\lambda_+^2}
\Biggl[
2\sin(2\alpha\lambda_-)\sin(2\alpha\lambda_+)
\Bigl(
1+(-1+2\tilde{g}_C^2)\cosh(4r)-2\tilde{g}_C^2\sinh(4r)
\Bigr)
\lambda_-\lambda_+}
\\[1ex]
&\qquad
+\cos^2(\alpha\lambda_-)\lambda_-^2
\Bigl[
4\sin^2(\alpha\lambda_+)
\Bigl(
(1+2\tilde{g}_C+2\tilde{g}_C^2)\cosh(4r)
-2\tilde{g}_C(1+\tilde{g}_C)\sinh(4r)
\Bigr)
+4\cos^2(\alpha\lambda_+)\lambda_+^2
\Bigr]
\\[1ex]
&\qquad
+4\sin^2(\alpha\lambda_-)
\Bigl[
(1+4\tilde{g}_C^2)\sin^2(\alpha\lambda_+)
+\cos^2(\alpha\lambda_+)
\Bigl(
(1-2\tilde{g}_C+2\tilde{g}_C^2)\cosh(4r)
-2(\tilde{g}_C-1)\tilde{g}_C\sinh(4r)
\Bigr)
\lambda_+^2
\Bigr]
\Biggr]
\\[2ex]
&\qquad
-
\sqrt{
-16+
\frac{1}{\lambda_-^4\lambda_+^4}
\Biggl[
2\sin(2\alpha\lambda_-)\sin(2\alpha\lambda_+)
\Bigl(
1+(-1+2\tilde{g}_C^2)\cosh(4r)-2\tilde{g}_C^2\sinh(4r)
\Bigr)
\lambda_-\lambda_+}
\\[1ex]
&\qquad\qquad
+\cos^2(\alpha\lambda_-)\lambda_-^2
\Bigl[
4\sin^2(\alpha\lambda_+)
\Bigl(
(1+2\tilde{g}_C+2\tilde{g}_C^2)\cosh(4r)
-2\tilde{g}_C(1+\tilde{g}_C)\sinh(4r)
\Bigr)
+4\cos^2(\alpha\lambda_+)\lambda_+^2
\Bigr]
\\[1ex]
&\qquad\qquad
+4\sin^2(\alpha\lambda_-)
\Bigl[
(1+4\tilde{g}_C^2)\sin^2(\alpha\lambda_+)
+\cos^2(\alpha\lambda_+)
\Bigl(
(1-2\tilde{g}_C+2\tilde{g}_C^2)\cosh(4r)
-2(\tilde{g}_C-1)\tilde{g}_C\sinh(4r)
\Bigr)
\lambda_+^2
\Bigr]
\Biggr]^2
\Biggr).
\label{eq:da}
\end{aligned}
\end{equation}

\begin{equation}
\begin{aligned}
&\tilde{d}_{\text{SM}}(\tilde{g}_C,\tilde{g}_D=0,\alpha,N,r,\theta=\pi)=
\\[1ex]
&\frac{1}{2}(1+2N)
\sqrt{
\Biggl(
\frac{1}{\lambda_-^2\lambda_+^2}
\Biggl[
2\sin(2\alpha\lambda_-)\sin(2\alpha\lambda_+)
\Bigl(
1+(-1+2\tilde{g}_C^2)\cosh(4r)+2\tilde{g}_C^2\sinh(4r)
\Bigr)
\lambda_-\lambda_+}
\\[1ex]
&\qquad
+\cos^2(\alpha\lambda_-)\lambda_-^2
\Bigl[
4\sin^2(\alpha\lambda_+)
\Bigl(
(1+2\tilde{g}_C+2\tilde{g}_C^2)\cosh(4r)
+2\tilde{g}_C(1+\tilde{g}_C)\sinh(4r)
\Bigr)
+4\cos^2(\alpha\lambda_+)\lambda_+^2
\Bigr]
\\[1ex]
&\qquad
+4\sin^2(\alpha\lambda_-)
\Bigl[
(1+4\tilde{g}_C^2)\sin^2(\alpha\lambda_+)
+\cos^2(\alpha\lambda_+)
\Bigl(
(1-2\tilde{g}_C+2\tilde{g}_C^2)\cosh(4r)
+2(\tilde{g}_C-1)\tilde{g}_C\sinh(4r)
\Bigr)
\lambda_+^2
\Bigr]
\Biggr]
\\[2ex]
&\qquad
-
\sqrt{
-16+
\frac{1}{\lambda_-^4\lambda_+^4}
\Biggl[
2\sin(2\alpha\lambda_-)\sin(2\alpha\lambda_+)
\Bigl(
1+(-1+2\tilde{g}_C^2)\cosh(4r)+2\tilde{g}_C^2\sinh(4r)
\Bigr)
\lambda_-\lambda_+}
\\[1ex]
&\qquad\qquad
+\cos^2(\alpha\lambda_-)\lambda_-^2
\Bigl[
4\sin^2(\alpha\lambda_+)
\Bigl(
(1+2\tilde{g}_C+2\tilde{g}_C^2)\cosh(4r)
+2\tilde{g}_C(1+\tilde{g}_C)\sinh(4r)
\Bigr)
+4\cos^2(\alpha\lambda_+)\lambda_+^2
\Bigr]
\\[1ex]
&\qquad\qquad
+4\sin^2(\alpha\lambda_-)
\Bigl[
(1+4\tilde{g}_C^2)\sin^2(\alpha\lambda_+)
+\cos^2(\alpha\lambda_+)
\Bigl(
(1-2\tilde{g}_C+2\tilde{g}_C^2)\cosh(4r)
+2(\tilde{g}_C-1)\tilde{g}_C\sinh(4r)
\Bigr)
\lambda_+^2
\Bigr]
\Biggr]^2
\Biggr).
\label{eq:db}
\end{aligned}
\end{equation}

\begin{equation}
\begin{aligned}
&\tilde{d}_{\text{TM}}(\tilde{g}_C,\tilde{g}_D=0,\alpha,N,r,\theta=0)=
\\[1ex]
&\frac{1}{\sqrt{2}}(1+2N)
\sqrt{
\Biggl(
\frac{1}{\lambda_-^{2}\lambda_+^{2}}
\Biggl[
\sin(2\alpha\lambda_-)\sin(2\alpha\lambda_+)
\Bigl(
-1 + 2\tilde{g}_C^2 + \cosh(4r) - 2\tilde{g}_C\sinh(4r)
\Bigr)
\lambda_-\lambda_+}
\\[1ex]
&\qquad
+2\sin^2(\alpha\lambda_-)
\Bigl[
\sin^2(\alpha\lambda_+)
\Bigl(
(1+4\tilde{g}_C^2)\cosh(4r)
-4\tilde{g}_C\sinh(4r)
\Bigr)
+(1-2\tilde{g}_C+2\tilde{g}_C^2)\cos^2(\alpha\lambda_+)\lambda_+^2
\Bigr]
\\[1ex]
&\qquad
+2\cos^2(\alpha\lambda_-)\lambda_-^2
\Bigl[
(1+2\tilde{g}_C+2\tilde{g}_C^2)\sin^2(\alpha\lambda_+)
+\cos^2(\alpha\lambda_+)\cosh(4r)\lambda_+^2
\Bigr]
\Biggr]
\\[2ex]
&\qquad
-
\sqrt{
-4+
\frac{1}{\lambda_-^{4}\lambda_+^{4}}
\Biggl[
\sin(2\alpha\lambda_-)\sin(2\alpha\lambda_+)
\Bigl(
-1 + 2\tilde{g}_C^2 + \cosh(4r) - 2\tilde{g}_C\sinh(4r)
\Bigr)
\lambda_-\lambda_+}
\\[1ex]
&\qquad\qquad
+2\sin^2(\alpha\lambda_-)
\Bigl[
\sin^2(\alpha\lambda_+)
\Bigl(
(1+4\tilde{g}_C^2)\cosh(4r)
-4\tilde{g}_C\sinh(4r)
\Bigr)
+(1-2\tilde{g}_C+2\tilde{g}_C^2)\cos^2(\alpha\lambda_+)\lambda_+^2
\Bigr]
\\[1ex]
&\qquad\qquad
+2\cos^2(\alpha\lambda_-)\lambda_-^2
\Bigl[
(1+2\tilde{g}_C+2\tilde{g}_C^2)\sin^2(\alpha\lambda_+)
+\cos^2(\alpha\lambda_+)\cosh(4r)\lambda_+^2
\Bigr]
\Biggr]^2
\Biggr).
\label{eq:dc}
\end{aligned}
\end{equation}

\begin{equation}
\begin{aligned}
&\tilde{d}_{\text{TM}}(\tilde{g}_C,\tilde{g}_D=0,\alpha,N,r,\theta=\pi)=
\\[1ex]
&\frac{1}{\sqrt{2}}(1+2N)
\sqrt{
\Biggl(
\frac{1}{\lambda_-^{2}\lambda_+^{2}}
\Biggl[
\sin(2\alpha\lambda_-)\sin(2\alpha\lambda_+)
\Bigl(
-1 + 2\tilde{g}_C^2 + \cosh(4r) + 2\tilde{g}_C\sinh(4r)
\Bigr)
\lambda_-\lambda_+}
\\[1ex]
&\qquad
+2\sin^2(\alpha\lambda_-)
\Bigl[
\sin^2(\alpha\lambda_+)
\Bigl(
(1+4\tilde{g}_C^2)\cosh(4r) + 4\tilde{g}_C\sinh(4r)
\Bigr)
+(1-2\tilde{g}_C+2\tilde{g}_C^2)\cos^2(\alpha\lambda_+)\lambda_+^2
\Bigr]
\\[1ex]
&\qquad
+2\cos^2(\alpha\lambda_-)\lambda_-^2
\Bigl[
(1+2\tilde{g}_C+2\tilde{g}_C^2)\sin^2(\alpha\lambda_+)
+\cos^2(\alpha\lambda_+)\cosh(4r)\lambda_+^2
\Bigr]
\Biggr]
\\[2ex]
&\qquad
-
\sqrt{
-4+
\frac{1}{\lambda_-^{4}\lambda_+^{4}}
\Biggl[
\sin(2\alpha\lambda_-)\sin(2\alpha\lambda_+)
\Bigl(
-1 + 2\tilde{g}_C^2 + \cosh(4r) + 2\tilde{g}_C\sinh(4r)
\Bigr)
\lambda_-\lambda_+}
\\[1ex]
&\qquad\qquad
+2\sin^2(\alpha\lambda_-)
\Bigl[
\sin^2(\alpha\lambda_+)
\Bigl(
(1+4\tilde{g}_C^2)\cosh(4r) + 4\tilde{g}_C\sinh(4r)
\Bigr)
+(1-2\tilde{g}_C+2\tilde{g}_C^2)\cos^2(\alpha\lambda_+)\lambda_+^2
\Bigr]
\\[1ex]
&\qquad\qquad
+2\cos^2(\alpha\lambda_-)\lambda_-^2
\Bigl[
(1+2\tilde{g}_C+2\tilde{g}_C^2)\sin^2(\alpha\lambda_+)
+\cos^2(\alpha\lambda_+)\cosh(4r)\lambda_+^2
\Bigr]
\Biggr]^2
\Biggr).
\label{eq:dd}
\end{aligned}
\end{equation}
\end{widetext}
where for the sake of clarity we have defined $\lambda_+=\sqrt{1+2\tilde{g}_C}$ and $\lambda_-=\sqrt{1-2\tilde{g}_C}$.
\newpage
\addcontentsline{toc}{section}{Bibliography}
\bibliography{biblio}

@article{Anupam1,
  title = {Relativistic effects on entangled single-electron traps},
  author = {Toro\ifmmode \check{s}\else \v{s}\fi{}, Marko and Andriolo, Patrick and Schut, Martine and Bose, Sougato and Mazumdar, Anupam},
  journal = {Phys. Rev. D},
  volume = {110},
  issue = {5},
  pages = {056031},
  numpages = {12},
  year = {2024},
  month = {Sep},
  publisher = {American Physical Society},
  doi = {10.1103/PhysRevD.110.056031},
  url = {https://link.aps.org/doi/10.1103/PhysRevD.110.056031}
}

@article{hc1,
  title = {Mass-energy and anomalous friction in quantum optics},
  author = {Sonnleitner, Matthias and Barnett, Stephen M.},
  journal = {Phys. Rev. A},
  volume = {98},
  issue = {4},
  pages = {042106},
  numpages = {12},
  year = {2018},
  month = {Oct},
  publisher = {American Physical Society},
  doi = {10.1103/PhysRevA.98.042106},
  url = {https://link.aps.org/doi/10.1103/PhysRevA.98.042106}
}

@article{hc2,
  title = {Post-Newtonian Hamiltonian description of an atom in a weak gravitational field},
  author = {Schwartz, Philip K. and Giulini, Domenico},
  journal = {Phys. Rev. A},
  volume = {100},
  issue = {5},
  pages = {052116},
  numpages = {16},
  year = {2019},
  month = {Nov},
  publisher = {American Physical Society},
  doi = {10.1103/PhysRevA.100.052116},
  url = {https://link.aps.org/doi/10.1103/PhysRevA.100.052116}
}

@book{hc3,
  author       = {V. B. Berestetskii and E. M. Lifshitz and L. P. Pitaevskii},
  title        = {Quantum Electrodynamics},
  series       = {Landau and Lifshitz Course of Theoretical Physics},
  volume       = {4},
  edition      = {2},
  year         = {1982},
  publisher    = {Elsevier},
  translator   = {J. B. Sykes and J. S. Bell}
}

@article{Covariance1,
  author    = {Mehdi Ahmadi and David Edward Bruschi and Carlos Sab{\'i}n and Gerardo Adesso and Ivette Fuentes},
  title     = {Relativistic Quantum Metrology: Exploiting relativity to improve quantum measurement technologies},
  journal   = {Scientific Reports},
  year      = {2014},
  volume    = {4},
  number    = {1},
  pages     = {4996},
  doi       = {10.1038/srep04996},
  url       = {https://doi.org/10.1038/srep04996},
  issn      = {2045-2322}
}

@article{Covariance2,
  title = {Quantum metrology for relativistic quantum fields},
  author = {Ahmadi, Mehdi and Bruschi, David Edward and Fuentes, Ivette},
  journal = {Phys. Rev. D},
  volume = {89},
  issue = {6},
  pages = {065028},
  numpages = {10},
  year = {2014},
  month = {Mar},
  publisher = {American Physical Society},
  doi = {10.1103/PhysRevD.89.065028},
  url = {https://link.aps.org/doi/10.1103/PhysRevD.89.065028}
}

@article{Canosa_2015,
doi = {10.1088/0953-4075/48/16/165501},
url = {https://doi.org/10.1088/0953-4075/48/16/165501},
year = {2015},
month = {jul},
publisher = {IOP Publishing},
volume = {48},
number = {16},
pages = {165501},
author = {Canosa, N and Mandal, Swapan and Rossignoli, R},
title = {Exact dynamics and squeezing in two harmonic modes coupled through angular momentum},
journal = {Journal of Physics B: Atomic, Molecular and Optical Physics}
}

@article{Olivares2012,
  author    = {S. Olivares},
  title     = {Quantum optics in the phase space},
  journal   = {The European Physical Journal Special Topics},
  year      = {2012},
  volume    = {203},
  number    = {1},
  pages     = {3--24},
  doi       = {10.1140/epjst/e2012-01532-4},
  url       = {https://doi.org/10.1140/epjst/e2012-01532-4},
  issn      = {1951-6401}
}

@article{Friis01012013,
author = {Nicolai Friis and Ivette Fuentes},
title = {Entanglement generation in relativistic quantum fields},
journal = {Journal of Modern Optics},
volume = {60},
number = {1},
pages = {22--27},
year = {2013},
publisher = {Taylor \& Francis},
doi = {10.1080/09500340.2012.712725},


URL = { 
    
        https://doi.org/10.1080/09500340.2012.712725
    
    

},
eprint = { 
    
        https://doi.org/10.1080/09500340.2012.712725
    
    

}

}

@article{Adesso_2007,
  title = {Continuous-variable entanglement sharing in noninertial frames},
  author = {Adesso, Gerardo and Fuentes-Schuller, Ivette and Ericsson, Marie},
  journal = {Phys. Rev. A},
  volume = {76},
  issue = {6},
  pages = {062112},
  numpages = {19},
  year = {2007},
  month = {Dec},
  publisher = {American Physical Society},
  doi = {10.1103/PhysRevA.76.062112},
  url = {https://link.aps.org/doi/10.1103/PhysRevA.76.062112}
}

@article{Adesso_2012,
doi = {10.1088/0264-9381/29/22/224002},
url = {https://doi.org/10.1088/0264-9381/29/22/224002},
year = {2012},
month = {oct},
publisher = {IOP Publishing},
volume = {29},
number = {22},
pages = {224002},
author = {Adesso, Gerardo and Ragy, Sammy and Girolami, Davide},
title = {Continuous variable methods in relativistic quantum information: characterization of quantum and classical correlations of scalar field modes in noninertial frames},
journal = {Classical and Quantum Gravity}}

@article{Illuminati,
doi = {10.1088/1751-8113/40/28/S01},
url = {https://doi.org/10.1088/1751-8113/40/28/S01},
year = {2007},
month = {jun},
publisher = {},
volume = {40},
number = {28},
pages = {7821},
author = {Adesso, Gerardo and Illuminati, Fabrizio},
title = {Entanglement in continuous-variable systems: recent advances and current perspectives},
journal = {Journal of Physics A: Mathematical and Theoretical}
}

@article{Weedbrook,
  title = {Gaussian quantum information},
  author = {Weedbrook, Christian and Pirandola, Stefano and Garc\'{\i}a-Patr\'on, Ra\'ul and Cerf, Nicolas J. and Ralph, Timothy C. and Shapiro, Jeffrey H. and Lloyd, Seth},
  journal = {Rev. Mod. Phys.},
  volume = {84},
  issue = {2},
  pages = {621--669},
  numpages = {0},
  year = {2012},
  month = {May},
  publisher = {American Physical Society},
  doi = {10.1103/RevModPhys.84.621},
  url = {https://link.aps.org/doi/10.1103/RevModPhys.84.621}
}

@article{Illuminati2,
doi = {10.1088/0953-4075/37/2/L02},
url = {https://doi.org/10.1088/0953-4075/37/2/L02},
year = {2003},
month = {dec},
publisher = {},
volume = {37},
number = {2},
pages = {L21},
author = {Alessio Serafini and Fabrizio Illuminati and Silvio De Siena},
title = {Symplectic invariants, entropic measures and correlations of Gaussian states},
journal = {Journal of Physics B: Atomic, Molecular and Optical Physics}
}

@article{Thermal1,
  title = {Scalable Quantum Processor with Trapped Electrons},
  author = {Ciaramicoli, G. and Marzoli, I. and Tombesi, P.},
  journal = {Phys. Rev. Lett.},
  volume = {91},
  issue = {1},
  pages = {017901},
  numpages = {4},
  year = {2003},
  month = {Jun},
  publisher = {American Physical Society},
  doi = {10.1103/PhysRevLett.91.017901},
  url = {https://link.aps.org/doi/10.1103/PhysRevLett.91.017901}
}

@phdthesis{Yu2025,
  author       = {Q. Yu},
  title        = {Quantum Information Processing with Trapped Electrons in Paul Traps},
  school       = {University of California, Berkeley},
  year         = {2025},
  type         = {PhD dissertation},
  url          = {https://escholarship.org/uc/item/56w664td},
  note         = {ProQuest ID: 32402220}
}

@article{Yang2015,
  author  = {Yang, Xihua and Xiao, Min},
  title   = {Electromagnetically Induced Entanglement},
  journal = {Scientific Reports},
  year    = {2015},
  volume  = {5},
  number  = {1},
  pages   = {13609},
  doi     = {10.1038/srep13609},
  url     = {https://doi.org/10.1038/srep13609},
  issn    = {2045-2322}
}

@article{qgem1,
  title = {Spin Entanglement Witness for Quantum Gravity},
  author = {Bose, Sougato and Mazumdar, Anupam and Morley, Gavin W. and Ulbricht, Hendrik and Toro\ifmmode \check{s}\else \v{s}\fi{}, Marko and Paternostro, Mauro and Geraci, Andrew A. and Barker, Peter F. and Kim, M. S. and Milburn, Gerard},
  journal = {Phys. Rev. Lett.},
  volume = {119},
  issue = {24},
  pages = {240401},
  numpages = {6},
  year = {2017},
  month = {Dec},
  publisher = {American Physical Society},
  doi = {10.1103/PhysRevLett.119.240401},
  url = {https://link.aps.org/doi/10.1103/PhysRevLett.119.240401}
}

@article{qgem2,
  title = {Gravitationally Induced Entanglement between Two Massive Particles is Sufficient Evidence of Quantum Effects in Gravity},
  author = {Marletto, C. and Vedral, V.},
  journal = {Phys. Rev. Lett.},
  volume = {119},
  issue = {24},
  pages = {240402},
  numpages = {5},
  year = {2017},
  month = {Dec},
  publisher = {American Physical Society},
  doi = {10.1103/PhysRevLett.119.240402},
  url = {https://link.aps.org/doi/10.1103/PhysRevLett.119.240402}
}

@article{darkmatter,
  title = {Entanglement based tomography to probe new macroscopic forces},
  author = {Barker, Peter F. and Bose, Sougato and Marshman, Ryan J. and Mazumdar, Anupam},
  journal = {Phys. Rev. D},
  volume = {106},
  issue = {4},
  pages = {L041901},
  numpages = {7},
  year = {2022},
  month = {Aug},
  publisher = {American Physical Society},
  doi = {10.1103/PhysRevD.106.L041901},
  url = {https://link.aps.org/doi/10.1103/PhysRevD.106.L041901}
}

@article{string,
  title = {Quantum entanglement of masses with nonlocal gravitational interaction},
  author = {Beckering Vinckers, Ulrich K. and de la Cruz-Dombriz, \'Alvaro and Mazumdar, Anupam},
  journal = {Phys. Rev. D},
  volume = {107},
  issue = {12},
  pages = {124036},
  numpages = {12},
  year = {2023},
  month = {Jun},
  publisher = {American Physical Society},
  doi = {10.1103/PhysRevD.107.124036},
  url = {https://link.aps.org/doi/10.1103/PhysRevD.107.124036}
}

@article{Massivegraviton,
  title = {Probing massless and massive gravitons via entanglement in a warped extra dimension},
  author = {Elahi, Shafaq Gulzar and Mazumdar, Anupam},
  journal = {Phys. Rev. D},
  volume = {108},
  issue = {3},
  pages = {035018},
  numpages = {14},
  year = {2023},
  month = {Aug},
  publisher = {American Physical Society},
  doi = {10.1103/PhysRevD.108.035018},
  url = {https://link.aps.org/doi/10.1103/PhysRevD.108.035018}
}

@article{MatterLight1,
  title = {Gravitational optomechanics: Photon-matter entanglement via graviton exchange},
  author = {Biswas, Dripto and Bose, Sougato and Mazumdar, Anupam and Toro\ifmmode \check{s}\else \v{s}\fi{}, Marko},
  journal = {Phys. Rev. D},
  volume = {108},
  issue = {6},
  pages = {064023},
  numpages = {14},
  year = {2023},
  month = {Sep},
  publisher = {American Physical Society},
  doi = {10.1103/PhysRevD.108.064023},
  url = {https://link.aps.org/doi/10.1103/PhysRevD.108.064023}
}

@article{MatterLight2,
  author    = {Pablo Guillermo Carmona Rufo and Anupam Mazumdar and Sougato Bose and Carlos Sab{\'\i}n},
  title     = {Digital quantum simulation of gravitational optomechanics with IBM quantum computers},
  journal   = {EPJ Quantum Technology},
  year      = {2024},
  volume    = {11},
  number    = {1},
  pages     = {31},
  doi       = {10.1140/epjqt/s40507-024-00242-0},
  url       = {https://doi.org/10.1140/epjqt/s40507-024-00242-0},
  issn      = {2196-0763}
}

@article{Weak2,
  title = {Distinguishing Jordan and Einstein frames in gravity through entanglement},
  author = {Chakraborty, Sumanta and Mazumdar, Anupam and Pradhan, Ritapriya},
  journal = {Phys. Rev. D},
  volume = {108},
  issue = {12},
  pages = {L121505},
  numpages = {6},
  year = {2023},
  month = {Dec},
  publisher = {American Physical Society},
  doi = {10.1103/PhysRevD.108.L121505},
  url = {https://link.aps.org/doi/10.1103/PhysRevD.108.L121505}
}

@article{Precision,
doi = {10.1088/0957-0233/1/2/001},
url = {https://doi.org/10.1088/0957-0233/1/2/001},
year = {1990},
month = {feb},
publisher = {},
volume = {1},
number = {2},
pages = {93},
author = {R C Thompson},
title = {Precision measurement aspects of ion traps},
journal = {Measurement Science and Technology}
}

@book{Earnshaw,
  title={Classical Electrodynamics},
  author={Jackson, J.D.},
  isbn={9780471309321},
  lccn={97046873},
  url={https://books.google.es/books?id=FOBBEAAAQBAJ},
  year={1998},
  publisher={Wiley}
}

@article{Metrology,
  author = {Rosenband, T. and Hume, D. B. and Schmidt, P. O. and Chou, C. W. and Brusch, A. and Lorini, L. and Oskay, W. H. and Drullinger, R. E. and Fortier, T. M. and Stalnaker, J. E. and Diddams, S. A. and Swann, W. C. and Newbury, N. R. and Itano, W. M. and Wineland, D. J. and Bergquist, J. C.},
  title = {Frequency ratio of Al$^{+}$ and Hg$^{+}$ single-ion optical clocks: metrology at the 17th decimal place},
  journal = {Science},
  volume = {319},
  number = {5871},
  pages = {1808--1812},
  year = {2008},
  doi = {10.1126/science.1154622}
}

@article{Paul1,
  title = {Electromagnetic traps for charged and neutral particles},
  author = {Paul, Wolfgang},
  journal = {Rev. Mod. Phys.},
  volume = {62},
  issue = {3},
  pages = {531--540},
  numpages = {0},
  year = {1990},
  month = {Jul},
  publisher = {American Physical Society},
  doi = {10.1103/RevModPhys.62.531},
  url = {https://link.aps.org/doi/10.1103/RevModPhys.62.531}
}

@article{Paul2,
  title = {Quantum dynamics of single trapped ions},
  author = {Leibfried, D. and Blatt, R. and Monroe, C. and Wineland, D.},
  journal = {Rev. Mod. Phys.},
  volume = {75},
  issue = {1},
  pages = {281--324},
  numpages = {0},
  year = {2003},
  month = {Mar},
  publisher = {American Physical Society},
  doi = {10.1103/RevModPhys.75.281},
  url = {https://link.aps.org/doi/10.1103/RevModPhys.75.281}
}

@article{start1,
  author = {Kastner, Marc},
  title = {Technology and the single electron},
  journal = {Nature},
  volume = {389},
  number = {6652},
  pages = {667--667},
  year = {1997},
  doi = {10.1038/39450},
  url = {https://doi.org/10.1038/39450}
}

@article{start2,
  author = {Blatt, Rainer and Wineland, David},
  title = {Entangled states of trapped atomic ions},
  journal = {Nature},
  volume = {453},
  number = {7198},
  pages = {1008--1015},
  year = {2008},
  doi = {10.1038/nature07125},
  url = {https://doi.org/10.1038/nature07125}
}

@article{start3,
  author = {Blinov, B. B. and Leibfried, D. and Monroe, C. and Wineland, D. J.},
  title = {Quantum Computing with Trapped Ion Hyperfine Qubits},
  journal = {Quantum Information Processing},
  volume = {3},
  number = {1},
  pages = {45--59},
  year = {2004},
  doi = {10.1007/s11128-004-9417-3},
  url = {https://doi.org/10.1007/s11128-004-9417-3}
}

@article{start4,
  title = {Observation of Entangled States of a Fully Controlled 20-Qubit System},
  author = {Friis, Nicolai and Marty, Oliver and Maier, Christine and Hempel, Cornelius and Holz\"apfel, Milan and Jurcevic, Petar and Plenio, Martin B. and Huber, Marcus and Roos, Christian and Blatt, Rainer and Lanyon, Ben},
  journal = {Phys. Rev. X},
  volume = {8},
  issue = {2},
  pages = {021012},
  numpages = {20},
  year = {2018},
  month = {Apr},
  publisher = {American Physical Society},
  doi = {10.1103/PhysRevX.8.021012},
  url = {https://link.aps.org/doi/10.1103/PhysRevX.8.021012}
}

@article{fundamental1,
doi = {10.1088/1367-2630/18/9/093050},
url = {https://doi.org/10.1088/1367-2630/18/9/093050},
year = {2016},
month = {sep},
publisher = {IOP Publishing},
volume = {18},
number = {9},
pages = {093050},
author = {Bushev, P A and Cole, J H and Sholokhov, D and Kukharchyk, N and Zych, M},
title = {Single electron relativistic clock interferometer},
journal = {New Journal of Physics}
}

@book{fundamental2,
  editor = {Quint, Wolfgang and Vogel, Manuel},
  title = {Fundamental Physics in Particle Traps},
  series = {Springer Tracts in Modern Physics},
  publisher = {Springer},
  year = {2014},
  doi = {10.1007/978-3-642-45201-7},
  isbn = {978-3-642-45201-7},
  url = {https://doi.org/10.1007/978-3-642-45201-7}
}

@article{fundamental3,
  title = {Trapped Electrons and Ions as Particle Detectors},
  author = {Carney, Daniel and H\"affner, Hartmut and Moore, David C. and Taylor, Jacob M.},
  journal = {Phys. Rev. Lett.},
  volume = {127},
  issue = {6},
  pages = {061804},
  numpages = {7},
  year = {2021},
  month = {Aug},
  publisher = {American Physical Society},
  doi = {10.1103/PhysRevLett.127.061804},
  url = {https://link.aps.org/doi/10.1103/PhysRevLett.127.061804}
}

@article{qc1,
  author = {Kielpinski, D. and Monroe, C. and Wineland, D. J.},
  title = {Architecture for a large-scale ion-trap quantum computer},
  journal = {Nature},
  volume = {417},
  number = {6890},
  pages = {709--711},
  year = {2002},
  doi = {10.1038/nature00784},
  url = {https://doi.org/10.1038/nature00784}
}

@article{qc2,
  title = {Demonstration of a Fundamental Quantum Logic Gate},
  author = {Monroe, C. and Meekhof, D. M. and King, B. E. and Itano, W. M. and Wineland, D. J.},
  journal = {Phys. Rev. Lett.},
  volume = {75},
  issue = {25},
  pages = {4714--4717},
  numpages = {0},
  year = {1995},
  month = {Dec},
  publisher = {American Physical Society},
  doi = {10.1103/PhysRevLett.75.4714},
  url = {https://link.aps.org/doi/10.1103/PhysRevLett.75.4714}
}

@article{qc3,
  title = {Feasibility study of quantum computing using trapped electrons},
  author = {Yu, Qian and Alonso, Alberto M. and Caminiti, Jackie and Beck, Kristin M. and Sutherland, R. Tyler and Leibfried, Dietrich and Rodriguez, Kayla J. and Dhital, Madhav and Hemmerling, Boerge and H\"affner, Hartmut},
  journal = {Phys. Rev. A},
  volume = {105},
  issue = {2},
  pages = {022420},
  numpages = {10},
  year = {2022},
  month = {Feb},
  publisher = {American Physical Society},
  doi = {10.1103/PhysRevA.105.022420},
  url = {https://link.aps.org/doi/10.1103/PhysRevA.105.022420}
}

@article{qc4,
doi = {10.1088/0953-4075/42/15/154010},
url = {https://doi.org/10.1088/0953-4075/42/15/154010},
year = {2009},
month = {jul},
publisher = {},
volume = {42},
number = {15},
pages = {154010},
author = {Marzoli, I and Tombesi, P and Ciaramicoli, G and Werth, G and Bushev, P and Stahl, S and Schmidt-Kaler, F and Hellwig, M and Henkel, C and Marx, G and Jex, I and Stachowska, E and Szawiola, G and Walaszyk, A},
title = {Experimental and theoretical challenges for the trapped electron quantum computer},
journal = {Journal of Physics B: Atomic, Molecular and Optical Physics}
}

@article{qc5,
  title = {Trapped electrons in vacuum for a scalable quantum processor},
  author = {Ciaramicoli, G. and Marzoli, I. and Tombesi, P.},
  journal = {Phys. Rev. A},
  volume = {70},
  issue = {3},
  pages = {032301},
  numpages = {16},
  year = {2004},
  month = {Sep},
  publisher = {American Physical Society},
  doi = {10.1103/PhysRevA.70.032301},
  url = {https://link.aps.org/doi/10.1103/PhysRevA.70.032301}
}

@article{qi1,
  title = {Monoelectron Oscillator},
  author = {Wineland, D. and Ekstrom, P. and Dehmelt, H.},
  journal = {Phys. Rev. Lett.},
  volume = {31},
  issue = {21},
  pages = {1279--1282},
  numpages = {0},
  year = {1973},
  month = {Nov},
  publisher = {American Physical Society},
  doi = {10.1103/PhysRevLett.31.1279},
  url = {https://link.aps.org/doi/10.1103/PhysRevLett.31.1279}
}

@article{qi2,
  title = {Geonium theory: Physics of a single electron or ion in a Penning trap},
  author = {Brown, Lowell S. and Gabrielse, Gerald},
  journal = {Rev. Mod. Phys.},
  volume = {58},
  issue = {1},
  pages = {233--311},
  numpages = {0},
  year = {1986},
  month = {Jan},
  publisher = {American Physical Society},
  doi = {10.1103/RevModPhys.58.233},
  url = {https://link.aps.org/doi/10.1103/RevModPhys.58.233}
}

@article{qi3,
  title = {Two-electron quantum dots as scalable qubits},
  author = {Jefferson, J. H. and Fearn, M. and Tipton, D. L. J. and Spiller, T. P.},
  journal = {Phys. Rev. A},
  volume = {66},
  issue = {4},
  pages = {042328},
  numpages = {11},
  year = {2002},
  month = {Oct},
  publisher = {American Physical Society},
  doi = {10.1103/PhysRevA.66.042328},
  url = {https://link.aps.org/doi/10.1103/PhysRevA.66.042328}
}

@article{qi4,
  author = {Yamahata, Gento and Nishiguchi, Katsuhiko and Fujiwara, Akira},
  title = {Gigahertz single-trap electron pumps in silicon},
  journal = {Nature Communications},
  volume = {5},
  pages = {5038},
  year = {2014},
  doi = {10.1038/ncomms6038},
  url = {https://doi.org/10.1038/ncomms6038}
}

@article{qi5,
  author = {Segal, Dvira and Shapiro, Moshe},
  title = {Nanoscale Paul Trapping of a Single Electron},
  journal = {Nano Letters},
  volume = {6},
  number = {8},
  pages = {1622--1626},
  year = {2006},
  publisher = {American Chemical Society},
  doi = {10.1021/nl060560h},
  url = {https://doi.org/10.1021/nl060560h},
  issn = {1530-6984}
}

@article{qi6,
doi = {10.1088/1367-2630/15/7/073017},
url = {https://doi.org/10.1088/1367-2630/15/7/073017},
year = {2013},
month = {jul},
publisher = {IOP Publishing},
volume = {15},
number = {7},
pages = {073017},
author = {Daniilidis, Nikos and Gorman, Dylan J and Tian, Lin and Häffner, Hartmut},
title = {Quantum information processing with trapped electrons and superconducting electronics},
journal = {New Journal of Physics}
}

@article{groups1,
  title = {Trapping Electrons in a Room-Temperature Microwave Paul Trap},
  author = {Matthiesen, Clemens and Yu, Qian and Guo, Jinen and Alonso, Alberto M. and H\"affner, Hartmut},
  journal = {Phys. Rev. X},
  volume = {11},
  issue = {1},
  pages = {011019},
  numpages = {12},
  year = {2021},
  month = {Jan},
  publisher = {American Physical Society},
  doi = {10.1103/PhysRevX.11.011019},
  url = {https://link.aps.org/doi/10.1103/PhysRevX.11.011019}
}

@article{groups2,
  title = {Spin readout of trapped electron qubits},
  author = {Peng, Pai and Matthiesen, Clemens and H\"affner, Hartmut},
  journal = {Phys. Rev. A},
  volume = {95},
  issue = {1},
  pages = {012312},
  numpages = {8},
  year = {2017},
  month = {Jan},
  publisher = {American Physical Society},
  doi = {10.1103/PhysRevA.95.012312},
  url = {https://link.aps.org/doi/10.1103/PhysRevA.95.012312}
}

@phdthesis{groups3,
  author       = {Günther Lientschnig},
  title        = {Quantum Information Processing with Trapped Electrons in Paul Traps},
  year         = {2002},
  type         = {PhD dissertation},
}

@article{groups4,
  title = {Scalable Quantum Processor with Trapped Electrons},
  author = {Ciaramicoli, G. and Marzoli, I. and Tombesi, P.},
  journal = {Phys. Rev. Lett.},
  volume = {91},
  issue = {1},
  pages = {017901},
  numpages = {4},
  year = {2003},
  month = {Jun},
  publisher = {American Physical Society},
  doi = {10.1103/PhysRevLett.91.017901},
  url = {https://link.aps.org/doi/10.1103/PhysRevLett.91.017901}
}

@article{Kumar2023continuousvariable,
  title = {Continuous-Variable Entanglement through Central Forces: Application to Gravity between Quantum Masses},
  author = {Kumar, Ankit and Krisnanda, Tanjung and Arumugam, Paramasivan and Paterek, Tomasz},
  journal = {Quantum},
  volume = {7},
  pages = {1008},
  year = {2023},
  month = may,
  doi = {10.22331/q-2023-05-15-1008},
  url = {https://doi.org/10.22331/q-2023-05-15-1008},
  issn = {2521-327X}
}

@article{Demarie_2018,
doi = {10.1088/1361-6404/aaaad0},
url = {https://doi.org/10.1088/1361-6404/aaaad0},
year = {2018},
month = {mar},
publisher = {IOP Publishing},
volume = {39},
number = {3},
pages = {035302},
author = {Demarie, Tommaso F},
title = {Pedagogical introduction to the entropy of entanglement for Gaussian states},
journal = {European Journal of Physics}
}

@article{death1,
  title = {Finite-Time Disentanglement Via Spontaneous Emission},
  author = {Yu, Ting and Eberly, J. H.},
  journal = {Phys. Rev. Lett.},
  volume = {93},
  issue = {14},
  pages = {140404},
  numpages = {4},
  year = {2004},
  month = {Sep},
  publisher = {American Physical Society},
  doi = {10.1103/PhysRevLett.93.140404},
  url = {https://link.aps.org/doi/10.1103/PhysRevLett.93.140404}
}

@article{death2,
  title = {Dark periods and revivals of entanglement in a two-qubit system},
  author = {Ficek, Z. and Tana\ifmmode \acute{s}\else \'{s}\fi{}, R.},
  journal = {Phys. Rev. A},
  volume = {74},
  issue = {2},
  pages = {024304},
  numpages = {4},
  year = {2006},
  month = {Aug},
  publisher = {American Physical Society},
  doi = {10.1103/PhysRevA.74.024304},
  url = {https://link.aps.org/doi/10.1103/PhysRevA.74.024304}
}

@article{death3,
  title = {Role of the Bell singlet state in the suppression of disentanglement},
  author = {Liu, Ru-Fen and Chen, Chia-Chu},
  journal = {Phys. Rev. A},
  volume = {74},
  issue = {2},
  pages = {024102},
  numpages = {4},
  year = {2006},
  month = {Aug},
  publisher = {American Physical Society},
  doi = {10.1103/PhysRevA.74.024102},
  url = {https://link.aps.org/doi/10.1103/PhysRevA.74.024102}
}

@article{death0,
title = {A study on the sudden death of entanglement},
journal = {Physics Letters A},
volume = {365},
number = {1},
pages = {44-48},
year = {2007},
issn = {0375-9601},
doi = {https://doi.org/10.1016/j.physleta.2006.12.049},
url = {https://www.sciencedirect.com/science/article/pii/S0375960106019980},
author = {H.T. Cui and K. Li and X.X. Yi}
}

@article{deathions,
title = {Sudden death and long-lived entanglement of two trapped ions},
journal = {Physics Letters A},
volume = {369},
number = {5},
pages = {372-376},
year = {2007},
issn = {0375-9601},
doi = {https://doi.org/10.1016/j.physleta.2007.05.003},
url = {https://www.sciencedirect.com/science/article/pii/S0375960107006974},
author = {Mahmoud Abdel-Aty and H. Moya-Cessa}
}

@article{LIGO,
  title = {Detection of 15 dB Squeezed States of Light and their Application for the Absolute Calibration of Photoelectric Quantum Efficiency},
  author = {Vahlbruch, Henning and Mehmet, Moritz and Danzmann, Karsten and Schnabel, Roman},
  journal = {Phys. Rev. Lett.},
  volume = {117},
  issue = {11},
  pages = {110801},
  numpages = {5},
  year = {2016},
  month = {Sep},
  publisher = {American Physical Society},
  doi = {10.1103/PhysRevLett.117.110801},
  url = {https://link.aps.org/doi/10.1103/PhysRevLett.117.110801}
}
\bibliographystyle{unsrt}    

\end{document}